\documentclass{aa}  
\usepackage{graphicx}
\usepackage{multirow}
\usepackage{txfonts}
\usepackage{lscape}
\begin{document}

   \title{Disc galaxies are still settling}
   
   \subtitle{The discovery of the smallest nuclear discs and their young stellar bars}

   \author{Camila de Sá-Freitas
          \inst{1}\thanks{\email{camila.desafreitas@eso.org}}
          \and Dimitri A. Gadotti\inst{2}
          \and Francesca Fragkoudi\inst{3}
          \and Lodovico Coccato\inst{1}
          \and Paula Coelho\inst{4}
          \and Adriana de Lorenzo-Cáceres\inst{5,6}
          \and Jesús Falcón-Barroso\inst{5,6}
          \and Tutku Kolcu\inst{7}
          \and Ignacio Martín-Navarro\inst{5,6}
          \and Jairo Mendez-Abreu\inst{5,6}
          \and Justus Neumann\inst{8}
          \and Patricia Sanchez Blazquez\inst{9,10}
          \and Miguel Querejeta\inst{11}
          \and Glenn van de Ven\inst{12}
         }

    \institute{European Southern Observatory, Karl-Schwarzschild-Str. 2, D-85748 Garching bei Muenchen, Germany
        \and Centre for Extragalactic Astronomy, Department of Physics, Durham University, South Road, Durham DH1 3LE, UK
        \and Institute for Computational Cosmology, Department of Physics, Durham University, South Road, Durham DH1 3LE, UK
        \and Universidade de São Paulo, Instituto de Astronomia, Geofísica e Ciências Atmosféricas, Rua do Matão 1226, 05508-090, São Paulo, SP, Brazil
        \and Instituto de Astrofísica de Canarias, Calle Vía Láctea s/n, 38205 La Laguna, Tenerife, Spain
        \and Departamento de Astrofísica, Universidad de La Laguna, 38200 La Laguna, Tenerife, Spain
        \and Astrophysics Research Institute, Liverpool John Moores University, IC2 Liverpool Science Park, 146 Brownlow Hill, L3 5RF Liverpool, UK
        \and Max Planck Institute for Astronomy, K\"onigstuhl 17, 69117, Germany
        \and Departamento de F\'{i}sica de la Tierra y Astrof\'{i}sica, Universidad Complutense de Madrid, E-28040 Madrid, Spain
        \and Instituto de F\'isica de Part\'iculas y del Cosmos (IPARCOS), Universidad Complutense de Madrid, E-28040 Madrid, Spain
        \and Observatorio Astronómico Nacional, C/Alfonso XII 3, Madrid 28014, Spain
        \and Department of Astrophysics, University of Vienna, Türkenschanzstraße 17, 1180 Wien, Austria
            }

   \date{Received September 15, 1996; accepted March 16, 1997}

 
  \abstract{When galactic discs settle and become massive enough, they are able to form stellar bars. These non-axisymmetric structures induce shocks in the gas, causing it to flow to the centre where nuclear structures, such as nuclear discs and rings, are formed. Previous theoretical and observational studies have hinted at the co-evolution of bars and nuclear discs, suggesting that nuclear discs grow "inside-out", thereby proposing that smaller discs live in younger bars. Nevertheless, it remains unclear how the bar and the nuclear structures form and evolve with time. The smallest nuclear discs discovered to date tend to be larger than $\sim200~\rm{pc}$, even though some theoretical studies find that when nuclear discs form they can be much smaller. Using MUSE archival data, we report for the first time two extragalactic nuclear discs with radius sizes below $100~\rm{pc}$. Additionally, our estimations reveal the youngest bars found to date. We estimate that the bars in these galaxies formed $4.50^{+1.60}_{-1.10}\rm{(sys)}^{+1.00}_{-0.75}\rm{(stat)}$ and $0.7^{+2.60}\rm{(sys)}^{+0.05}_{-0.05}\rm{(stat)}~\rm{Gyr}$ ago, for NGC\,289 and NGC\,1566, respectively. This suggests that at least some disc galaxies in the Local Universe may still be dynamically settling. By adding these results to previous findings in the literature, we retrieve a stronger correlation between nuclear disc size and bar length and we derive a tentative exponential growth scenario for nuclear discs. }



   \maketitle
%

\section{The realm of nuclear discs}


A large number of disc galaxies display an elongated structure, named a bar, in different redshifts. The fraction of barred galaxies increases with time (e.g., \citealp{sheth2008evolution}; \citealp{melvin2014galaxy}) and, for the Local Universe, reaches values of $30-70\%$, depending on mass cut and detection methods (e.g., \citealp{eskridge2000frequency}; \citealp{menendez2007near}; \citealp{barazza2008bars}; \citealp{aguerri2009population}; \citealp{nair2010fraction}; \citealp{masters2011galaxy}; \citealp{buta2015classical}; \citealp{erwin2018dependence}). 

The presence of the bar is an indication that the host galaxy is dynamically `settled', that is, the disc is self-gravitating, with differential rotation, and rotationally supported with relatively low-velocity dispersion (e.g., \citealp{kraljic2012two} and references therein). Additionally, the host galaxy is undergoing secular evolutionary processes driven by the bar. Among different aspects, bars are responsible for the redistribution of angular momentum (e.g., \citealp{lynden1972generating}; \citealp{combes1985spiral}; \citealp{athanassoula2003angular}; \citealp{sheth2005secular}; \citealp{di2013signatures}; \citealp{halle2015quantifying}; \citealp{fragkoudi2017bars}) and the creation of central substructures such as nuclear discs (e.g., \citealp{athanassoula1992morphology}, \citeyear{Athanassoula1992b}; \citealp{munoz2004inner}; \citealp{athanassoula2005nature}; \citealp{coelho2011bars}; \citealp{ellison2011impact}; \citealp{cole2014formation}; \citealp{emsellem2015interplay}; \citealp{fragkoudi2016close}; \citealp{seo2019effects}; \citealp{gadotti2020kinematic}; \citealp{baba2020age}), often named ``pseudo-bulges''. More precisely, the non-axisymmetric potential of the bar causes the gas in the main disc to shock and lose angular momentum, funneling inwards. The gas is halted in the central region of the galaxy with high rotational velocities, where it forms stars and gives rise to nuclear rings and/or discs. Nevertheless, \cite{cameron2010bars} show that $19\%$ of nuclear discs can be found in bar-less galaxies, suggesting that other formation mechanisms for these structures are also possible, such as gas inflows due to spiral arms or tidal interactions. One other possibility to explain nuclear discs in bar-less galaxies is the eventual destruction of the bar that primarily formed the nuclear structure. In fact, some early simulations that accounted for gas dynamics, such as \cite{bournaud2005lifetime}, find that bars could be recurrent structures. That is, they could be destroyed and renewed multiple times. However, more recent works predominantly find that bars are in fact long-lived structures in the absence of a major merger (e.g., \citealp{kraljic2012two}; \citealp{gadotti2015muse}; \citealp{perez2017observational}; \citealp{de2019clocking}; \citealp{rosas2020buildup}; \citealp{fragkoudi2020chemodynamics}, \citeyear{fragkoudi2021revisiting}; \citealp{de2023new}). In conclusion, once the galaxy presents a non-axisymmetric potential, most likely due to the presence of the bar, nuclear structures are formed by gas inflows.

Many galaxies in the Local Universe, including the Milky Way (e.g., \citealp{sormani2020jeans}, \citeyear{sormani2022self}), host nuclear discs and/or rings (e.g., \citealp{comeron2010ainur}; \citealp{sheth2010spitzer}; \citealp{gadotti2015muse}; \citealp{erwin2015composite}), which can vary in properties such as size, star formation rate (SFR), and gas and dust content. Considering an SDSS sample of $\sim1000$ galaxies and performing 2D image decompositions, \cite{gadotti2009structural} found that $32\%$ of disc galaxies with photometric bulges actually host a nuclear disc. The Atlas of Images of NUclear Rings (AINUR -- \citealp{comeron2010ainur}) shows that $20\%$ of the disc galaxies in the Local Universe host a star-forming nuclear ring. The Time Inference with MUSE in Extragalactic Rings survey (TIMER -- \citealp{gadotti2019time}) finds that for a sample of $21$ massive, strongly barred galaxies, morphologically selected as hosting nuclear rings, at least $19$ clearly host a rapidly rotating nuclear disc. Nevertheless, it is not clear how common these structures are for different morphologies, masses, and redshifts, and how accurate the different detection methods are.

Nuclear discs, which in the past were also referred to as ``pseudo-bulges'' (e.g. \citealp{kormendy2004secular}), can be differentiated from ``classical bulges'' by using photometry since they display exponential surface density profiles, characteristic of discs (e.g., \citealp{gadotti2020kinematic} and references therein). For that reason, they also have been called ``discy-bulge'' and other denominations, to differentiate these structures from classical dynamically-hot bulges (\citealp{athanassoula2005nature}). Even though these structures can be identified through photometry, results from the TIMER survey (\citealp{gadotti2019time}, \citeyear{gadotti2020kinematic}) argue that the chances of misclassification of nuclear discs can be high when the physical spatial resolution is not suitable, and the best way to find and characterize these structures is through high spatial resolution integral field spectroscopy, with the derivation of the spatial distributions of stellar kinematics and population properties. In agreement, \cite{mendez2018morpho} find no correlation between photometric and kinematic properties of bulges in a sample of $28$ lenticular galaxies from the Calar Alto Legacy Integral Field Area survey (CALIFA, \citealp{sanchez2012califa}). 

By carefully measuring the kinematic properties of nuclear discs, \cite{gadotti2020kinematic} found that the nuclear disc kinematic size is well correlated with the bar length, (in qualitative accordance with \citealp{shlosman1989bars}; \citealp{knapen2005structure}; \citealp{comeron2010ainur}), where, typically, longer bars host larger nuclear discs. The kinematic size is defined by the place in which the radial profile of stellar velocity over velocity dispersion ($V/\sigma$) is maximum. Additionally, some recent simulations suggest that as the bar grows longer, the nuclear disc also increases in size (e.g., \citealp{seo2019effects}). This indicates a possible co-evolution between the bar and the nuclear disc. This co-evolution can be explained by the place in which the gas stops moving inwards and forms the nuclear disc. Even though the exact location in which it happens is unclear, some works suggest it can be associated with bar properties, directly or indirectly. Early work indicates that the gas moving inwards stops at the Inner Linbald Resonance (ILR) and forms the nuclear disc (e.g., \citealp{athanassoula1992morphology}, \citeyear{Athanassoula1992b}). As the bar grows and evolves, the ILR moves outwards, building the nuclear disc inside-out. On the other hand, some works have suggested that the nuclear disc size is related to the residual angular momentum of the original gas  (e.g., \citealp{kim2012gaseous}; \citealp{seo2019effects}). As the bar grows longer, it reaches the outer regions of the galaxy, where the gas has higher angular momentum. In this scenario, gas brought inwards from the outer parts of the galaxies would have higher residual angular momentum, and stop funneling inwards earlier, also building the nuclear disc inside-out. In both scenarios described, we expect the nuclear disc evolution to be linked with the bar evolution, where the nuclear disc is built inside-out (e.g., \citealp{bittner2020inside}, \citealp{de2023new}). However, it remains not clear if bars grow with time and if the co-evolution with nuclear discs is real since nuclear discs can also grow independently of the bar (e.g., \citealp{athanassoula1992existence}). Lastly, different works suggest that nuclear rings are the outer rim of the nuclear disc, which is currently forming stars (e.g., \citealp{cole2014formation}; \citealp{bittner2020inside}). \cite{gadotti2020kinematic} and \cite{bittner2020inside} demonstrated that nuclear discs and nuclear rings have the same kinematic properties and should not be differentiated. 

Considering the inside-out growth scenario for the nuclear discs, one expects them to form small and increase in size with time (e.g., \citealp{seo2019effects}; \citealp{bittner2020inside}; \citealp{de2023new}). Additionally, since the nuclear disc size could be linked to bar properties, one can expect that recently-formed and small bars host small nuclear discs. If bars grow and if the co-evolution between them and nuclear discs is real, as the bar grows longer, the nuclear disc should grow as well. In fact, \cite{seo2019effects} find for a Milky-Way-like simulated galaxy that the nuclear disc can form as small as $40~\rm{pc}$ and grow with time. 

Even though stellar nuclear discs with sizes of a few dozen parsecs have been found in early-type galaxies (e.g., \citealp{ledo2010census}; \citealp{sarzi2016nuclear}; \citealp{corsini2016young}), the smallest kinematically confirmed nuclear discs reported so far tend to be larger than $\sim200~\rm{pc}$ (e.g., \citealp{gadotti2020kinematic}). Additionally, the few galaxies with measured bar ages are older than $7~\rm{Gyr}$ (e.g., \citealp{gadotti2015muse}; \citealp{perez2017observational}; \citealp{de2019clocking}; \citealp{de2023new}). It is therefore unclear if nuclear discs form with typical sizes of $\sim200~\rm{pc}$, or if they can form with smaller sizes and grow in time. It is also unclear if most barred galaxies will have old bars, or whether disc galaxies can still be in the process of forming their stellar bars. Using MUSE ESO archival data\footnote{http://archive.eso.org/scienceportal/home}, we aim to start answering these questions by reporting the discovery {of the smallest kinematically confirmed nuclear discs} to date, to the best of our knowledge, together with a characterisation of their properties. Moreover, we follow the methodology presented in \cite{de2023new} to estimate bar ages and investigate the properties of the bars hosting such small nuclear discs.

The structure of this paper is organized as follows: In Section~\ref{sec_SampleDataDescr} we describe the data and the characteristics of the galaxies in our study, NGC\,289 and NGC\,1566. In Section~\ref{sec_Methodology} we describe the expected characteristics of nuclear discs formed by bars and how one can estimate the bar formation epoch considering the star formation histories (SFHs) of bar-built structures. In Section~\ref{sec_Results} we present our results on the presence of small nuclear discs on both galaxies, with radius sizes below $100~\rm{pc}$, and what are the respectively estimated bar ages. In Section~\ref{sec_Discussion} we discuss the implications of our findings on galaxy secular evolution and the formation and evolution of nuclear discs and their host bars. We summarize our work in Section~\ref{sec_Summary}.

\section{Sample and data description }
\label{sec_SampleDataDescr}

In this section, we describe the galaxies hosting these small nuclear discs, NGC\,289 and NGC\,1566, together with the data description and observing program details. 

NGC\,289 is a weak barred spiral galaxy (T-type $4$ -- \citealp{sheth2010spitzer}) with the presence of rings (e.g., \citealp{1991rc3..book.....D}; \citealp{munoz2013impact}; \citealp{buta2015classical}), with stellar mass measurements varying between $3.2\times10^{10}~\rm{M}_\odot$ (\citealp{lopez2022exploring}) and $4\times10^{10}~\rm{M}_\odot$ (\citealp{sheth2010spitzer}, \citealp{munoz2015spitzer}), and inclination of $43^\circ$\footnote{http://leda.univ-lyon1.fr}. Considering the redshift-independent distance measurements distribution from the NASA/IPAC Extragalactic Database (NED\footnote{https://ned.ipac.caltech.edu}), we derive a median distance of $18$ Mpc. At that distance, the measured deprojected bar size of $18.4~\pm~0.4$'' from \cite{munoz2013impact} corresponds to $1.62~\pm~0.35$ kpc. Lastly, NGC\,289 has an interacting companion, the dwarf galaxy Arp\,1981 (e.g., \citealp{bendo2004nuclear}). We use ESO archive data\footnotemark[1] from the {MUSE Atlas of Disks program} (MAD -- \citealp{erroz2019muse}), PI: Carollo, M. C., program ID 096.B-0309, using the MUSE Wide Field Mode. The galaxy was observed on the 15$^{th}$ of October 2015, for a total integration time of 2400 seconds, with a point spread function (PSF) with full width at half-maximum (FWHM) of $0.6''$. More details regarding the galaxy, observation, and calibration can be found in \cite{erroz2019muse}.

NGC\,1566 is classified as a weakly barred galaxy (T-type $4$ -- \citealp{sheth2010spitzer}; \citealp{1991rc3..book.....D}), with rings -- nuclear and outer --  and spiral arms. The galaxy has stellar mass measurements between $3.8\times10^{10}~\mathrm{M}_\odot$ (\citealp{sheth2010spitzer}, \citealp{munoz2015spitzer}) and $6\times10^{10}~\mathrm{M}_\odot$ (\citealp{leroy2021phangs}), with an inclination of $32^\circ$ (\citealp{salo2015spitzer}), and it is at a median distance\footnotemark[2] of $7.3$ Mpc. At that distance, the measured deprojected bar size of $40.5\pm2.5$'' from \cite{munoz2013impact} corresponds to $1.4\pm0.1$ kpc. Lastly, NGC\,1566 belongs to the Dorado group and has a dwarf elliptical companion, NGC\,1581 (e.g., \citealp{kendall2015spiral}). We use ESO archive data\footnotemark[1] from the MAD program (\citealp{erroz2019muse}), PI: Carollo, M. C., program ID 0100.B-0116, using the MUSE Wide Field Mode with adaptive optics. The galaxy was observed on the 23$^{rd}$ of October 2017, for a total integration time of 3600 seconds, with a PSF FWHM of $1.0''$. More details regarding the galaxy, observation, and calibration can be found in \cite{erroz2019muse}.



\section{Analysis and methodology}
\label{sec_Methodology}
\subsection{Finding nuclear discs}
\label{sec_findingNuclearDiscs}

We expect most of the nuclear discs to form by gas infall due to the onset of a non-axisymmetric potential, such as the one produced by stellar bars. Within this scenario, there are common properties that one can expect nuclear discs to present.

Firstly, the stars formed by the gas will form the stellar nuclear disc, which will present higher rotational velocities than the stars already present in the central region of the galaxy. In addition, since the nuclear disc is a rotationally-supported structure, we expect low values of velocity dispersion. Once the nuclear disc forms, we have at least two structures co-existing: a cold, rapidly rotating system (the nuclear disc) and a more slowly rotating system of stars that were already present (the main disc). Considering they had different formation histories, epochs, and time scales, each structure rotates independently, that is, they have different dynamical properties. Since the light from the nuclear region carries information about both these structures, the absorption lines will not be perfect Gaussians, but display deviations. We can measure these deviations considering the Gaussian-Hermite higher-order moments $h_3$ and $h_4$ (\citealp{van1993new}), which measure asymmetric and symmetric deviations, respectively. A negative $h_3$ indicates an excess of stars rotating slower than the average system velocity, while a positive $h_3$ indicates the opposite, an excess of stars rotating faster than the average velocity. This explains why in the presence of a fast-rotating nuclear disc, there is an anti-correlation between stellar velocity and $h_3$: the region in which the nuclear disc is fastly approaching the observer (blue-shifted velocities), there is also the main disc approaching us slower, hence negatives values of $h_3$. The opposite is also true. On the other hand, a positive $h_4$ indicates the presence of two rotating systems with different velocity dispersion, generating a pointy Gaussian distribution. For more details about the Gaussian-Hermite higher-order moments, see fig.\,3 on \cite{gadotti2005vertical}. In summary, the expected kinematic properties for the presence of a nuclear disc are (\textit{i}) an increase in rotational stellar velocity, which is the line-of-sight velocity corrected for inclination, (\textit{ii}) a drop in stellar velocity dispersion, (\textit{iii}) an anti-correlation between stellar velocity and $h_3$, and (\textit{iv}) an increase in $h_4$ (e.g., \citealp{gadotti2020kinematic}).

The same formation scenario also predicts mean stellar population characteristics. Since the nuclear disc is formed by a gas inflow that only takes place once the bar potential is in place, its stars are expected to be younger than the stars from the main disc in the same region. Additionally, the gas brought inwards by the bar is likely to be already metal-enriched when forming the stars of the nuclear disc. Hence, the metallicity ([M/H]) of the nuclear disc is expected to increase and be higher than the surroundings. Nevertheless, depending on the metallicity gradient of the galaxy, the origin of the gas, and the star formation history of the nuclear disc, one can find different metallicity behaviors within the nuclear disc (e.g., \citealp{bittner2020inside}). Lastly, since the nuclear disc evolution is due to long secular evolution processes, it is slowly built with continuous star formation. Because of that, depending on the strength of ongoing star formation, we expect the nuclear disc to present lower $\alpha$-enhancement values ([$\alpha$/Fe]) than the surroundings -- at least for most nuclear discs. In summary, among the stellar population properties we expect the nuclear disc to present when compared to its surroundings are (\textit{v}) younger median stellar ages, (\textit{vi}) higher [M/H], and (\textit{vii}) lower [$\alpha$/Fe] (e.g., \citealp{cole2014formation}; \citealp{bittner2020inside}). Hence, to unmistakably identify the presence of nuclear discs in galaxies and their origin, one has to derive kinematic and stellar population properties from datacubes.

To derive the kinematic and stellar population properties of the galaxies in our sample, we use the \texttt{Galaxy IFU Spectroscopy Tool} (GIST -- \citealp{bittner2019gist}). GIST is a module-based pipeline that allows us to derive physical properties from fully reduced datacubes. To ensure consistency with previous works, we followed the analysis described in \cite{gadotti2020kinematic}, \cite{bittner2020inside} and \cite{de2023new}, in two independent runs.

\begin{figure*}
    \centering
    \includegraphics[width=\linewidth]{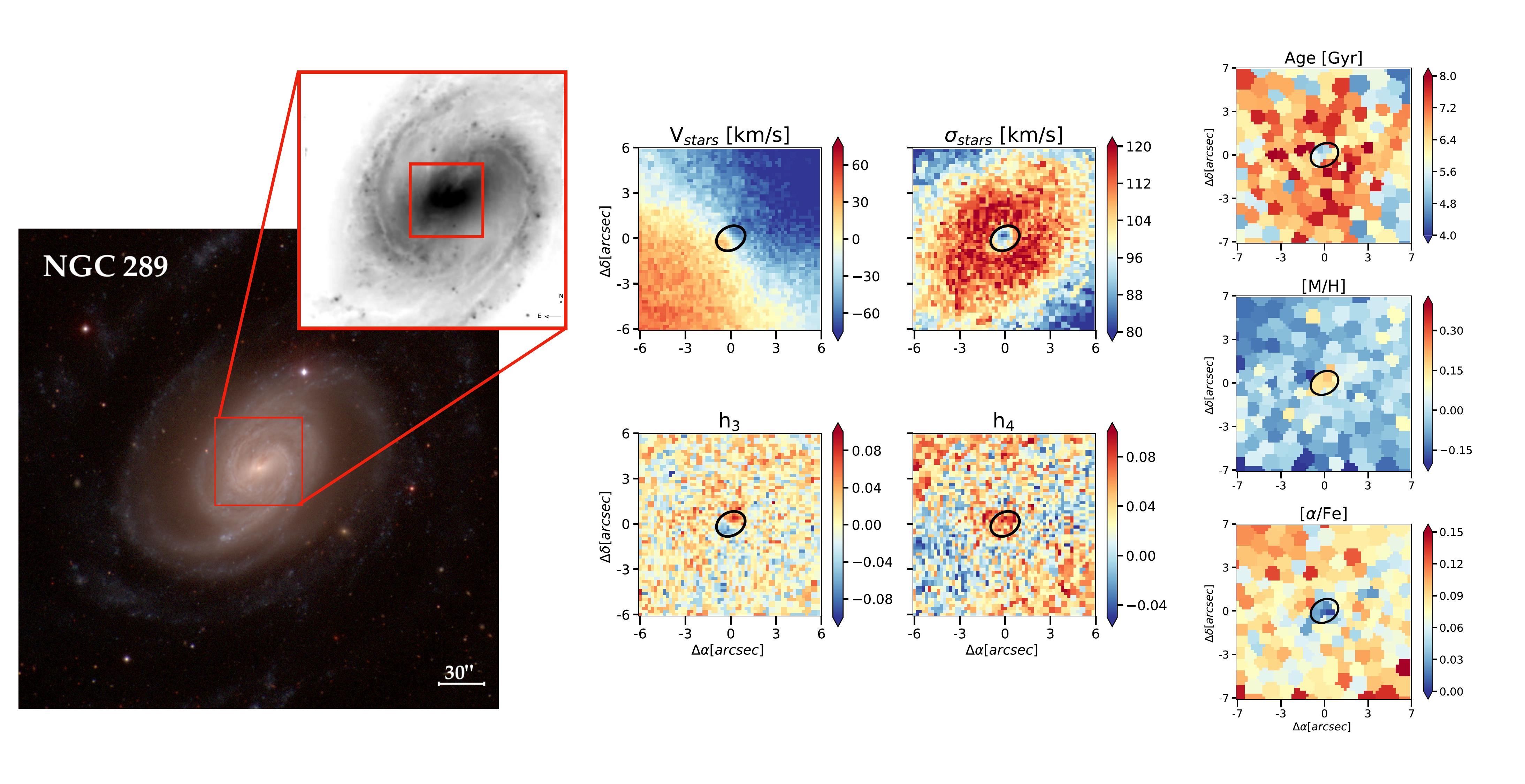}
    \caption{\textit{\textbf{NGC\,289 data and derived maps}}. On the left, we display the colour composites of NGC\,289 from the Carnegie-Irvine Galaxy Survey (CGS -- \citealp{ho2011carnegie}) together with the black and white image from MUSE ESO archival data (MAD -- \citealp{erroz2019muse}). We highlight the central region from which we derive the kinematic and stellar population maps. On the right, we display the seven spatial maps with derived kinematic and stellar population properties: stellar velocity (V$_{stars}$), stellar velocity dispersion ($\sigma_{stars}$), the Gauss-Hermite higher-order moments $h_3$ and $h_4$ (\citealp{van1993new}), mean age, metallicity ([M/H]), and $\alpha$ elements enhancement ([$\alpha$/Fe]). Together with the spatial maps, we display the limit of the nuclear disc in a black solid ellipse. We find a nuclear disc with a radius size of $90$ pc. Within the limits of the ellipse, one can notice all the expected properties of a nuclear disc: increase in stellar velocity, decrease in stellar velocity dispersion, anti-correlation between $h_3$ and the stellar velocity, increase in $h_4$, decrease in mean ages, increase in [M/H], and decrease in [$\alpha$/Fe].}
    \label{fig_NGC289}
\end{figure*}

In the first run, we aim to derive the kinematic properties of the galaxy. Firstly, GIST employs an unregularized run of \texttt{pPXF} (\citealp{cappellari2004parametric}; \citealp{cappellari2012ppxf}), considering the wavelength range between $4800-8950~$\r{A}. The data is binned following the Voronoi binning procedure (\citealp{cappellari2003adaptive}) to achieve a signal-to-noise of $40$. Additionally, we include a low-order multiplicative Legendre polynomial to account for differences between the observed spectra and the shape of the continuum templates. From this run, we retrieve spatial maps of stellar velocity, stellar velocity dispersion, $h_3$, and $h_4$. 

In the second run, we aim to derive the stellar population properties of the galaxy. We repeat the first step of the unregularized \texttt{pPXF} run, but considering a wavelength range of $4800-5800~$\r{A} and Voronoi-binning our sample to achieve a signal-to-noise of $100$. This choice is due to the fact that a higher signal-to-noise is more reliable when it comes to retrieving stellar population properties and avoiding spurious results between adjacent bins, as demonstrated in \cite{bittner2020inside}. Next, GIST employs \texttt{pyGandALF}, which is a \texttt{python} version of the \texttt{Gas and Absorption Line Fitting} (\texttt{gandALF} -- \citealp{sarzi2006sauron}; \citealp{falcon2006sauron}). This step consists of the modeling and removal of emission lines as Gaussians, resulting in the emission-subtracted spectra. In the last step, GIST employs a regularized run of \texttt{pPXF} in the emission-subtracted spectra, fitting different templates of stellar populations and enabling us to derive mean properties. We consider the MILES simple stellar population models library (\citealp{vazdekis2015evolutionary}), with [M/H] values between $-1$ and $+0.4$, ages between $0.03$ and $14$ Gyr, and [$\alpha$/Fe] enhancements of $+0.0$ and $+0.4$. We normalize the MILES templates for each mean flux, deriving light-weighted properties. In addition, since both metallicity and velocity dispersion can be responsible for broadening the absorption lines (e.g., \citealp{sanchez2011star}), we fixed the stellar kinematics from the unregularized \texttt{pPXF} run. Lastly, we use the regularization error value of $0.15$ (\citealp{bittner2020inside}) and apply an $8$th order multiplicative Legendre polynomial -- to account for possible extinction and continuum mismatches between the templates and the observed spectra. From the second run, we retrieve spatial maps of light-weighted mean values of stellar age, [M/H], and [$\alpha$/Fe]. Lastly, we would like the reader to keep in mind that for that specific wavelength, the dominant $\alpha$ element is magnesium (Mg) and our analysis refer to it.



\begin{figure*}
    \centering
    \includegraphics[width=\linewidth]{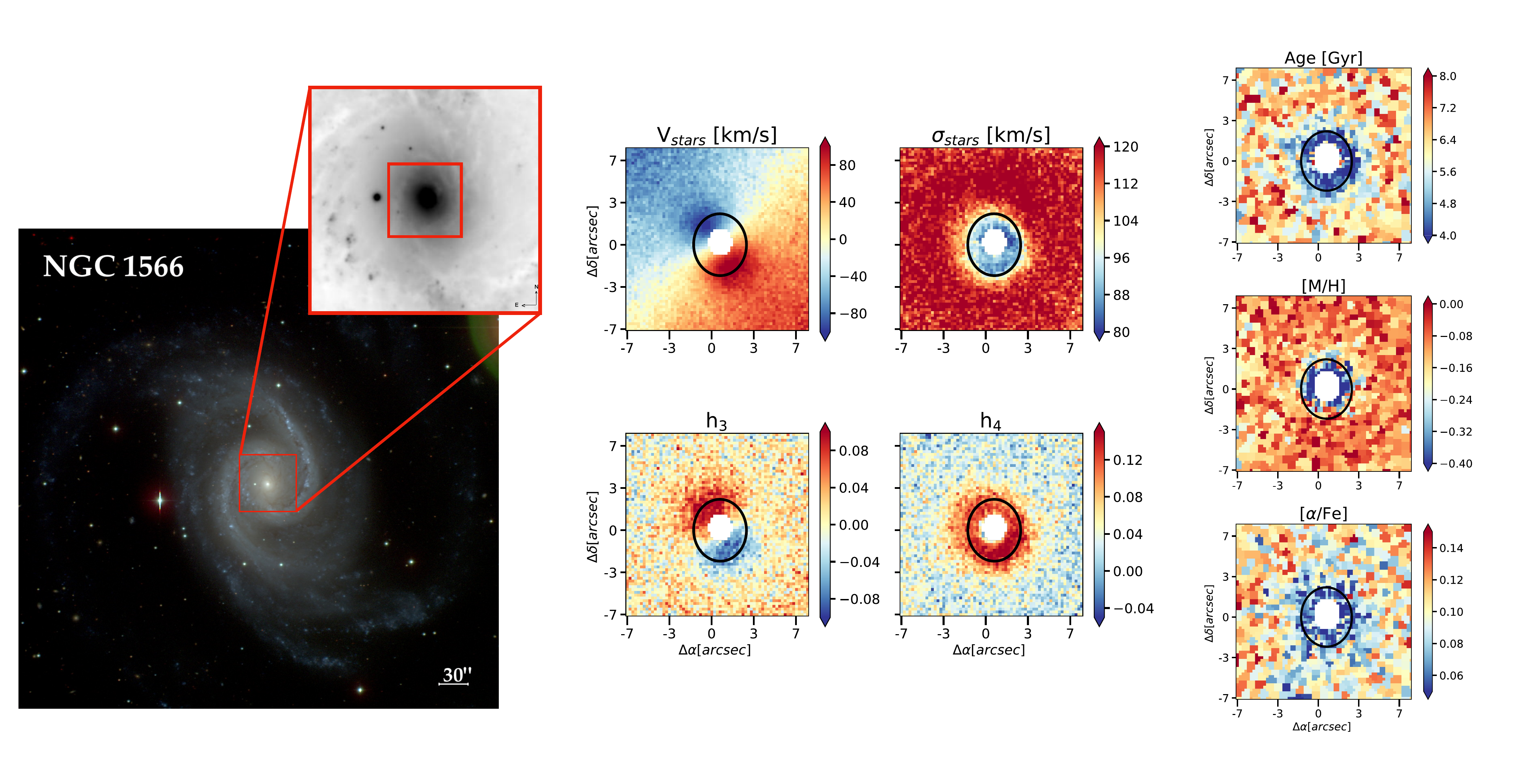}
    \caption{\textit{\textbf{NGC\,1566 data and derived maps}}. Same as Fig. \ref{fig_NGC289}. With the spatial maps, we display the limit of the nuclear disc measured considering the peak in the $V/\sigma$ radial profile, in a black solid ellipse. We find a nuclear disc with a radius size of $77$ pc. Within the limits of the ellipse, one can notice most of the properties expected for a nuclear disc: increase in stellar velocity, decrease in stellar velocity dispersion, anti-correlation between $h_3$ and the stellar velocity, increase in $h_4$, decrease in mean ages, and decrease in [$\alpha$/Fe]. The only property that differs from expected is the [M/H], which also decreases. This behavior can be related to the original properties of the infalling gas. Lastly, we mask the central region which presented strong emission lines, characteristic of AGN.}
    \label{fig_NGC1566}
\end{figure*}

\subsection{Estimating bar ages using nuclear discs}
\label{sec_EstimatingBarAgesMethod}

Numerical simulations have shown that when a bar forms, a nuclear disc forms within $10^8~\rm{yr}$, which is relatively short compared to the bar lifetime (which is of the order of a few $10^9~\rm{yr}$ -- \citealp{athanassoula1992morphology}, \citeyear{Athanassoula1992b}; \citealp{emsellem2015interplay}; \citealp{seo2019effects}; \citealp{baba2020age}). Considering that, one can derive the bar formation epoch measuring the ages of the stars in the nuclear disc. However, deriving such properties is not a trivial task, since the observed light from the nuclear disc also carries tangled information from stars that were already present when it formed -- that is, the main disc. With that in mind, we developed a methodology to disentangle the independent information from the nuclear disc and the main disc and, subsequently, estimate the time of the bar formation. For more details on the methodology, we refer the reader to \cite{de2023new}. In what follows we briefly summarize the different steps of the method: \textbf{first}, we convolve and shift all spectra in the datacube, ensuring the same velocity dispersion and velocity zero for all spaxels; \textbf{second}, we mask all spaxels classified as AGN using the BPT classification with amplitude over noise (AON) above 20 (\citealp{baldwin1981classification}); \textbf{third}, we select a ring region around the nuclear disc to derive the spectrum of the underlying main disc, hereafter denominated \textit{representative ring/spectrum} -- for this sample of small nuclear discs, we placed the representative ring region at $1.2$'' from the nuclear disc radius; \textbf{fourth}, using the representative spectrum and assuming an exponential light profile, we model the main disc datacube --  we use disc scale-lenghts values derived in \cite{salo2015spitzer}; \textbf{fifth}, we subtract the main disc from the original datacube -- shifted to velocity zero and convolved to maximum velocity dispersion --  and consider the difference as the light from the nuclear disc isolated; as an extra step, we collapsed each datacube into a average spectrum;  \textbf{lastly}, we employ GIST as described in Section~\ref{sec_findingNuclearDiscs} for the second run, deriving mean stellar populations for each collapsed spectrum (MUSE original, main disc, and nuclear disc).

During the fit of the emission-subtracted spectra, \texttt{pPXF} estimates different weights for different simple stellar populations (SSPs), differing in age, [M/H], and [$\alpha$/Fe]. These weights represent the fraction of the light due to the different SSPs. Considering the different weights for different SSPs, we are able to build light-weighted non-parametric SFHs for each collapsed spectrum (MUSE original, nuclear disc, and main disc). Finally, to convert the SFHs from light- to mass-weighted, we apply the mass-to-light ratios\footnote{http://research.iac.es/proyecto/miles/pages/predicted-masses-and-photometric-observables-based-on-photometric-libraries.php} predictions from the MILES models (\citealp{vazdekis2015evolutionary}), considering the BaSTI isochrones (\citealp{pietrinferni2004large}, \citeyear{pietrinferni2006large}, \citeyear{pietrinferni2009large}, \citeyear{pietrinferni2013basti}), converting luminosity into mass. The mentioned mass-to-light ratios assume a Kroupa revised IMF (\citealp{kroupa2001variation}) with [$\alpha$/Fe] enhancements of $+0.0$ and $+0.4$. Additionally, the ratios account for both stellar and remnants masses and depend on age, [M/H], and  [$\alpha$/Fe] that best describes the observed spectra. As a result, we can derive independent mass-weighted SFHs for the nuclear disc and main disc. 

Finally, we consider that shortly after the bar forms the stellar mass built by the nuclear disc increases above the stellar mass built by the main disc and, therefore, the ratio between the nuclear disc and the main disc rises above $1$, with a positive slope towards younger ages. This takes into account that it is possible to have residuals of old stellar populations in the nuclear disc, which is expected since the representative spectrum might not be as old as the underlying main disc. For more details on some of the tests carried out to test our methodology, as well as its caveats, we refer the reader to \cite{de2023new}.


%


\section{Results}
\label{sec_Results}
\subsection{Evidence of small nuclear discs}
\label{sec_evidenceSmallNuclearDisc}

\begin{figure*}
\centering
\includegraphics[width=0.7\linewidth]{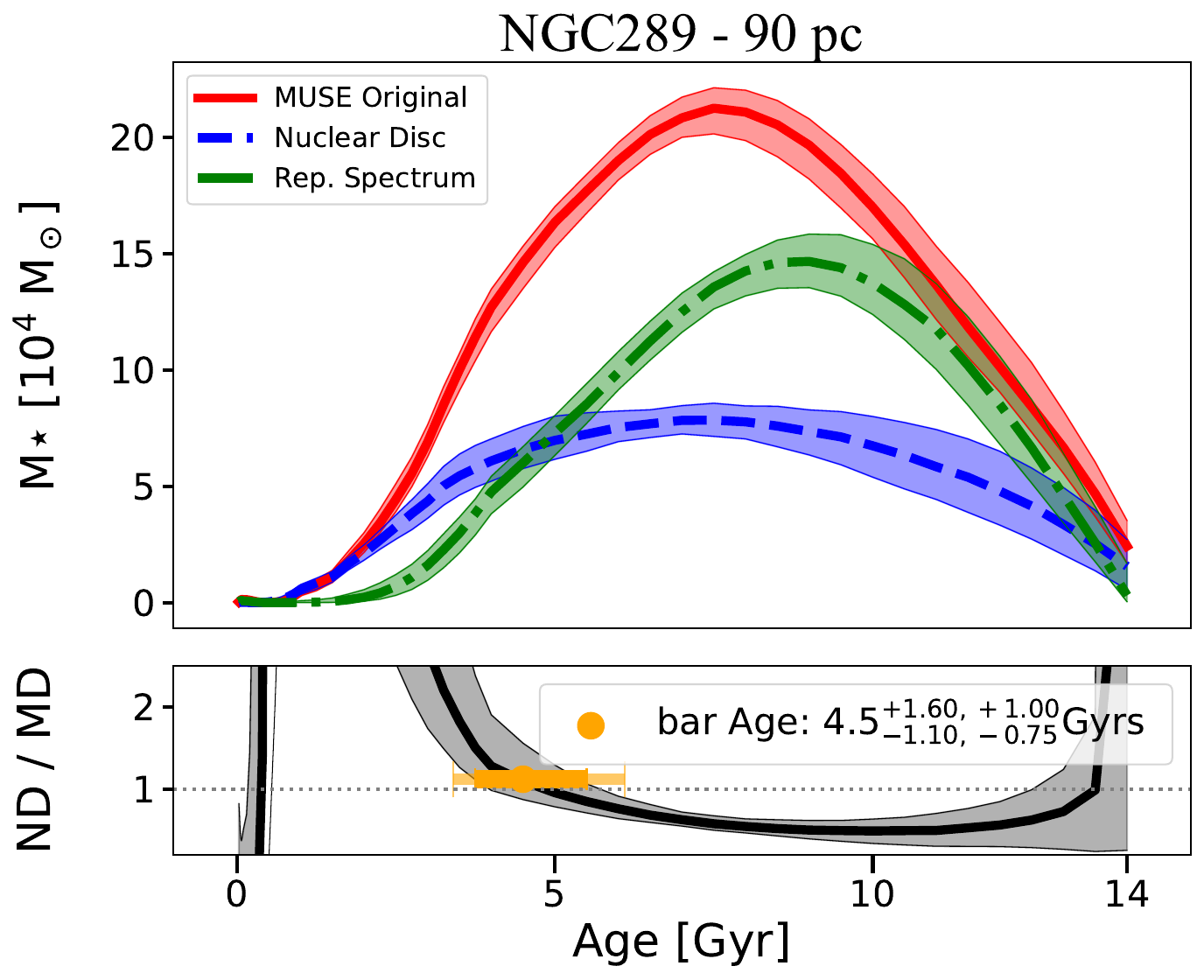}
    \caption{\textit{\textbf{NGC\,289 bar age measurement}}. On the top panel, we display star formation histories -- stellar mass built over time -- of the original data (red-solid line), the modeled main disc (green-dot-dashed line), and the nuclear disc isolated (blue-dashed line). With each SFH, we display the results from the $100$ MC runs (shaded regions), considering the $1^{st}$ and $9^{th}$ quantiles. On the bottom panel, we display the ratio between the nuclear disc and the main disc SFHs as a function of time (black-solid line), with the range of values from the $100$ MC runs (gray-shaded region). We consider the moment of bar formation when ND/MD $> 1$ towards younger ages. This moment is highlighted by the orange dot and marks an age of $4.5^{+1.60}_{-1.10}(\rm{sys})^{+1.00}_{-0.75}(\rm{stat})$ Gyr. Further discussion of the measurement of the presented errors can be found in Sect. \ref{sec_timingBarFormation}.}
    \label{fig_barAge_NGC289}
\end{figure*}

\begin{figure*}
\centering
\includegraphics[width=\linewidth]{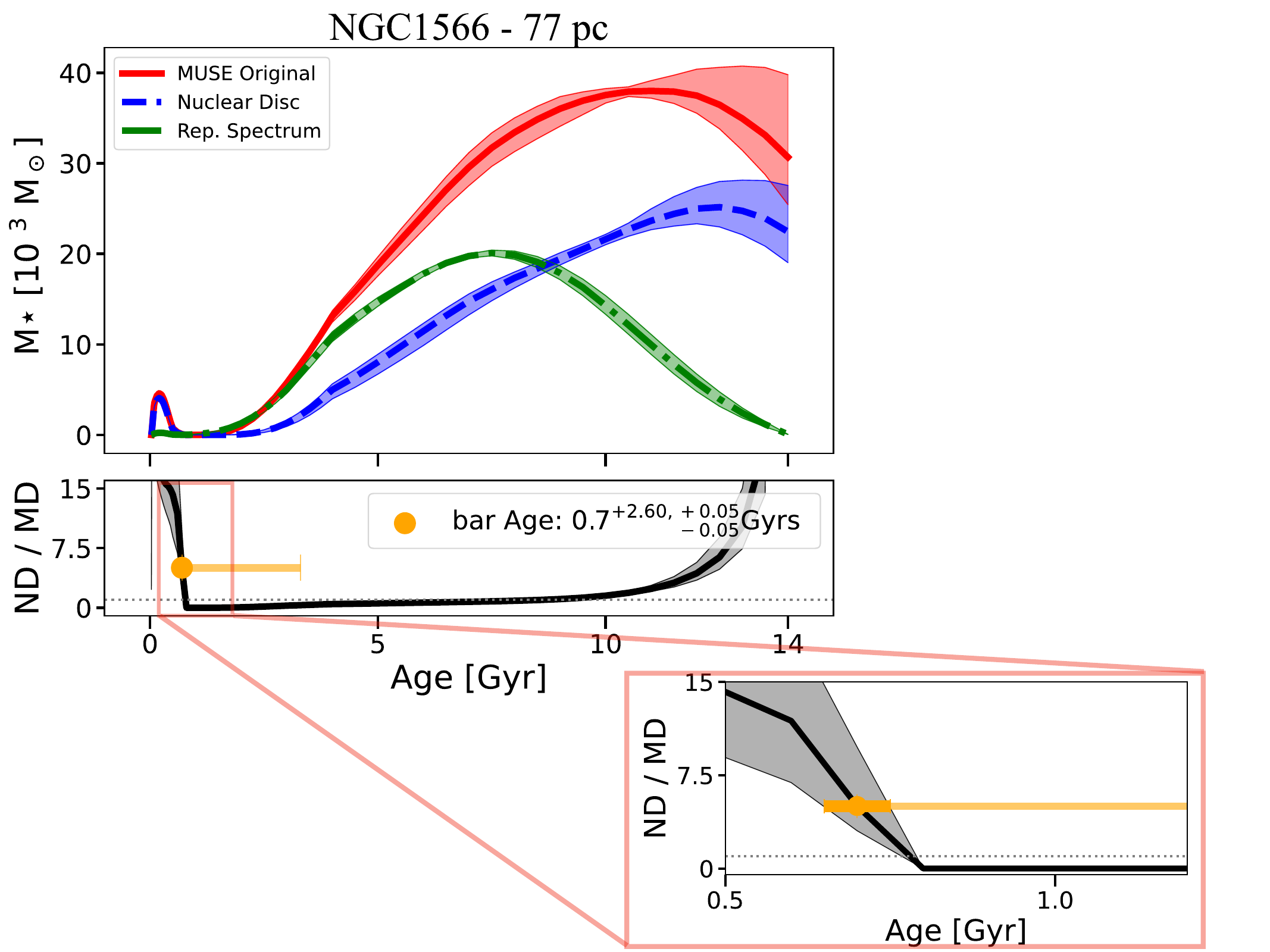}
    \caption{\textit{\textbf{NGC\,1566 bar age measurement}}. Same as Fig. \ref{fig_barAge_NGC289}. The criterion of ND/MD $> 1$ is highlighted by the orange dot and marks an age of $0.7^{+2.60}(\rm{sys})^{+0.05}_{-0.05}(\rm{stat})$ Gyr for the bar hosted by NGC\,1566. Additionally, we display a zoom-in of a region of the bottom panel, highlighting the variations due to the $100$ MC runs (gray-shaded regions).}
    \label{fig_barAge_NGC1566}
\end{figure*}

In this Section, we describe our results on the presence of nuclear discs on NGC\,289 and NGC\,1566. Figures \ref{fig_NGC289} and \ref{fig_NGC1566} display, for each galaxy, the object coloured-image from the Carnegie-Irvine Galaxy Survey (CGS - \citealp{ho2011carnegie}), the MUSE field of view, and the spatial maps of the kinematic and stellar population properties. The individual analysis and further discussion are below.

\textbf{\textit{NGC\,289}} -- Since the spatial resolution of the data is limited, we can not derive a meaningful radial profile of $V/\sigma$ within the nuclear disc, nor identify the position of its maximum value. Because of this, we determine the radius size of the nuclear disc visually, considering only the spatial maps in Fig. \ref{fig_NGC289}. We find the nuclear disc radius to be $1$'', which corresponds to a physical radius size of $90$ pc. This radius is very consistent across all seven assessed maps. Considering the $1^{st}$ and $3^{rd}$ quartiles of the distribution of distance measurements from NED, the nuclear disc radius size error is $90^{+16}_{-8}$ pc. Despite the apparent small size of the nuclear disc, the expected characteristics are still clear. Following the kinematic maps, we found an increase in the stellar velocity and a drop in the stellar velocity dispersion. Furthermore, considering the Gauss-Hermite higher-order moments $h_3$ and $h_4$, we identify an anti-correlation with $V$ for the former and an increase for the latter. All these kinematic properties indicate the presence of a second independent rotationally-supported structure, i.e., the nuclear disc. Additionally, the stellar population properties agree with the scenario that the nuclear disc was formed by gas inflow following the formation of the bar. The nuclear disc presents younger average ages than the surrounding regions, an increase in the [M/H], and a decrease in [$\alpha$/Fe] enhancements. We would like to stress how the powerful resolution of MUSE allows us to identify nuclear discs even in extreme cases such as the one of NGC\,289.

\textbf{\textit{NGC\,1566}} -- the results for NGC\,1566 are presented in Fig.~\ref{fig_NGC1566}. We mask the central region due to the presence of broad emission lines, characteristic of AGNs. Considering the $V/\sigma$ radial profile, we find a nuclear disc with an apparent radius of $2.2$'' corresponding to a physical radius size of $77$ pc. Considering the $1^{st}$ and $3^{rd}$ quartiles of the distribution of distance measurements from NED, the nuclear disc size error is $77^{+47}_{-2}$ pc. As NGC\,289, NGC\,1566 displays most of the characteristics of a young nuclear disc, when compared to the main disc: increase in stellar velocity rotation, decrease in stellar velocity dispersion, $h_3-V$ anti-correlation, increase in $h_4$, younger average stellar ages, and increase in [$\alpha$/Fe] enhancements. The only unexpected characteristic is the decrease in [M/H], where the opposite is expected for most bar-built nuclear discs. Nevertheless, \cite{bittner2020inside} found the same trend for $3$ nuclear discs and $8$ nuclear rings in a sample of $17$ galaxies. This behavior can be related to the original properties of the infalling gas.

In summary, the two galaxies present the kinematic characteristics of a nuclear disc and most of the expected stellar population properties. Even though the [M/H] values of NGC\,1566 are not necessarily consistent with most of the nuclear disc characteristics (e.g., \citealp{bittner2020inside}), the two nuclear discs display younger stellar ages in comparison to the surroundings, which is expected for bar-driven gas inflow. Lastly, it is worth it highlighting that the nuclear disc radius sizes of NGC\,289 and NGC\,1566 may not be consistent with each other, since we did not derive them following the same methodology. As noticeable in Fig. \ref{fig_NGC1566}, the characteristic kinematic radius size, based on the peak of $V/\sigma$, can be underestimating the nuclear disc size of NGC\,1566 when compared to NGC\,289.


\subsection{Timing bar formation}
\label{sec_timingBarFormation}

In this Section, we describe the measured ages for the bars hosted by NGC\,289 and NGC\,1566 following the methodology presented at \cite{de2023new}. In Figs. \ref{fig_barAge_NGC289} and \ref{fig_barAge_NGC1566} we present the Star Formation Histories -- stellar mass built over time -- for the original data, the nuclear disc, and the main disc for NGC\,289 and NGC\,1566, respectively. We also display the ratio between the stellar mass of the nuclear disc and the main disc for every given SSP age, with a highlight on the bar age. We constrain possible errors in our results, considering mainly two sources: data statistical errors and methodology systematic errors. We measured the data statistical error by performing $100$ Monte Carlo runs in the collapsed data of each datacube -- MUSE original, nuclear disc, and main disc. We consider the noise at each wavelength to sample a random distribution of fluxes, creating 100 artificial spectra. Following that, we repeat the methodology from Sec. \ref{sec_EstimatingBarAgesMethod} for each of the $300$ artificial datacubes, deriving a distribution of SFHs and bar ages, which we consider as the statistical error.

On the other hand, the measurement of systematic errors consists of quantifying how the configuration of our methodology can affect the final bar age. This includes different galactocentric distances of the representative ring, the light profile assumed to model the main disc (exponential or flat), the measured age for collapsed and non-collapsed versions of the datacubes, and the adopted regularization value in the final \texttt{pPXF} fit. Each different configuration results in a somewhat different bar age. We consider the difference from our main configuration to each test as a systematic error value. To quantify the final systematic error, we add all systematic errors in quadrature. In this work, we vary the position of the representative ring, and the assumed model profile (exponential or flat), and consider the collapsed and non-collapsed configurations. For the regularization value, we assume the error of $0.5$~Gyr depending on the chosen regularization value, found in \cite{de2023new}.

\renewcommand{\arraystretch}{1.4}
\begin{table*}[]
\centering
\begin{tabular}{cccp{0.20\linewidth}}
\textbf{Galaxy} & \textbf{Stellar mass (M$_\odot$)}  & \textbf{Reference} & \textbf{Method} \\
\hline
\hline
\multirow{3}{*}{NGC\,289}  & $4.3\times10^{10}$ & \cite{sheth2010spitzer} &  $3.6\mu$m  \\
&  $4.0\times10^{10}$ & \cite{lopez2022exploring} & SSP analysis \\
&  $4.6\times10^{10}$ & This work & SSP analysis and extrapolation of exponential disc \\
\hline
\multirow{3}{*}{NGC\,1433} & $2.0\times10^{10}$ & \cite{sheth2010spitzer} & $3.6\mu$m  \\
& $7.4\times10^{10}$ & \cite{leroy2021phangs} & $3.4\mu$m \\
& $2.62\times10^{10}$ & \cite{de2023new} & SSP analysis and extrapolation of exponential disc \\
\hline
\multirow{3}{*}{NGC\,1566} & $3.8\times10^{10}$ & \cite{sheth2010spitzer} & $3.6\mu$m\\
& $6.2\times10^{10}$ & \cite{leroy2021phangs} & $3.4\mu$m \\
& $8.3\times10^{10}$ &  This work & SSP analysis and extrapolation of exponential disc \\
\hline
\multirow{2}{*}{NGC\,4371} & $3.2\times10^{10}$ & \cite{sheth2010spitzer} & $3.6\mu$m \\
& $6.3\times10^{10}$ & \cite{gallo2010amuse} & $g_0 ~\rm{and}~ z_0$ bands \\
\hline
Milky Way & $6.1\pm1.14\times10^{10}$  & \cite{licquia2015improved}  & Hierarchical Bayesian combination of previous measurements from the literature  \\
\end{tabular}
\caption{Total stellar masses for the galaxies considered in Fig. \ref{fig_TIMER}, as derived in different studies and with different methods, as indicated.}
\label{tab_stellarMasses}
\end{table*}

\textbf{\textit{NGC\,289}} -- We find an age of $4.50$ Gyr for the bar hosted by NGC\,289 (see Fig. \ref{fig_barAge_NGC289}). Additionally, from the Monte Carlo runs, we measure a statistical error of $^{+1.00}_{-0.75}$ Gyr, considering the $1^{st}$ and $9^{th}$ quantiles of the distributions of SFHs. Compared to the statistical error we find for the bar in NGC\,1433 ($^{+0.2}_{-0.5}$ -- \citealp{de2023new}), NGC\,289 errors are $1.5-5$ times larger. This is expected due to the fact that the nuclear disc present in NGC\,1433 is much larger and better resolved. In that sense, the original data occupies more spaxels when compared to NGC\,289. Thus, once we collapse the datacube, we achieve a signal-to-noise close to $2000$ for NGC\,1433. On the other hand, the collapsed spectrum of NGC\,289 has a signal-to-noise of around $200$. Additionally, we constrained a systematic error of $+1.6$ on the bar age by varying the configurations on our methodology -- the position of the representative ring, modeled main disc light profile, running the analysis on a spaxel by spaxel basis rather than collapsing the datacube, and regularization. Lastly, \cite{de2023new} demonstrated that placing the representative ring closer to the nuclear disc can result in younger bar ages. More specifically, we found a systematic error of $1.1~$Gyr younger for bar ages. Since the data on NGC\,289 intrinsically has a low physical spatial resolution, we originally placed the representative ring the closest allowed by observational constraints, that is, $1.2$'' of distance determined by the seeing. Due to that, we can not explore the systematic error of placing the representative ring closer to the nuclear disc and opted to adopt the systematic error of $1.1$ Gyr younger for NGC\,289 as well. In summary, we find that NGC\,289 hosts a bar with an age of $4.50^{+1.60}_{-1.10}\rm{(sys)}^{+1.00}_{-0.75}\rm{(stat)}$ Gyr.

\textbf{\textit{NGC\,1566}} -- We find an age of $0.70$ Gyr for the bar hosted by NGC\,1566 (see Fig. \ref{fig_barAge_NGC1566}). For this galaxy, when applying our methodology we find an excess of old stellar populations in our nuclear disc. As discussed in \cite{de2023new}, this is likely due to the negative age gradient in the galaxy. Our method of obtaining the bar age is robust against biases introduced due to this old stellar population residual. From the Monte Carlo runs, we measure a statistical error of $\pm0.05$ Gyr, considering the $1^{st}$ and $9^{th}$ quantiles, considerably smaller than NGC\,289. The low statistical error for NGC\,1566 is mainly due to two facts: (\textit{i}) the signal to noise achieved by the collapsed datacube is over $1000$ and (\textit{ii}) the particular shape of the SFHs and the sudden peak in young ages is similar in the $100$ MC runs. Nevertheless, we also present a zoomed-in region in Fig. \ref{fig_barAge_NGC1566} to highlight the differences from the $100$ MC runs. In addition, we find systematic errors of ${+2.60}$ Gyr. The systematic errors are due to the differences between the collapsed and non-collapsed results and the regularization error adopted from \cite{de2023new}. Because NGC\,1566 is a nearby galaxy ($7.3~\rm{Mpc}$) and the observations were carried out with adaptative optics, the resolution is sufficient to test how varying the distance of the representative ring will affect our results, which is one of our main systematic uncertainties. However, the bar age retrieved for different representative rings distances remains the same, that is, $0.70$ Gyr. In summary, we find a bar age of $0.70^{+2.60}\rm{(sys)}^{+0.05}_{-0.05}\rm{(stat)}$ Gyr for NGC\,1566.

\subsection{Integrating the SFHs -- a consistency check}
\label{sec_Mass}

Assuming our decomposition of the central light in the two discs is correct and, taking into account their relative brightness, we can calculate their contribution to the mass integrating their SFH (see also \citealp{de2023new}). For NGC\,289 we measure $6.4\times10^7~\rm{M}_\odot$ and $4.4\times10^7~\rm{M}_\odot$ for the main and nuclear disc, respectively. In that sense, the recently formed nuclear disc accounts for $\sim41\%$ of the total mass budget within the central $90~\rm{pc}$. On the other hand, for NGC\,1566 we measure the main disc mass of $2.9\times10^7~\rm{M}_\odot$ and the nuclear disc mass of  $3.8\times10^7~\rm{M}_\odot$, where the nuclear disc accounts for $\sim56\%$ of the central mass budget within $77~\rm{pc}$. The measured masses for the nuclear discs are in good agreement with the findings in \cite{seo2019effects}, in which the authors find masses for recently formed nuclear discs of $4\times10^7~\rm{M}_\odot$ for Milky-Way-like galaxies, with stellar masses of $4.5-5\times10^{10}\rm{M}_\odot$. 

Furthermore, considering the surface mass density of the main disc, we are able to extrapolate the results above to estimate the total stellar mass of the galaxy assuming an exponential function following:

\begin{equation}
    \rm{M}_\star = 2\pi\int_0^{\infty}\Sigma(\rm{r})\rm{dr} = 2\pi\Sigma_0 h^2,
\end{equation}

\noindent where $\Sigma_0$ is the central surface density and h is the disc scale-length -- we consider $1.7$ kpc and $2.6$ kpc for NGC\,289 and NGC\,1566, respectively (\citealp{salo2015spitzer}). We find extrapolated total stellar masses of $4.6\times10^{10}~\rm{M}_\odot$, for NGC\,289, and $8.3\times10^{10}~\rm{M}_\odot$, for NGC\,1566. With these estimates for the total stellar mass of the galaxy, we can compare it to measurements that apply different methods, to do a consistency check of our structure disentanglement. In Table \ref{tab_stellarMasses} we summarize the different stellar masses measured for different galaxies, including NGC\,289 and NGC\,1566. We find extrapolated masses close to the literature, especially in the case of NGC\,289. On the other hand, the extrapolated mass for NGC\,1566 is larger than the ones measured by the S$^4$G (\citealp{sheth2010spitzer}) and PHANGS (\citealp{leroy2021phangs}) teams. Nevertheless, the values from both works also vary greatly, demonstrating that measuring the stellar mass content is not a trivial task.



\section{Small nuclear discs and young bars in the context of secular evolution}
\label{sec_Discussion}

In this Section, we discuss our results on the smallest nuclear discs reported and what insights they bring for galaxy secular evolution. We would like to stress that the results achieved by this work were only possible due to the incredible resolving power of state-of-the-art IFUs. Our results illustrate how we can uncover relatively compact structures, their kinematics, and their stellar population properties.


\subsection{The smallest nuclear discs discovered -- what does this tell us?}
\label{subSec_SmallestNuclearDisc}

\begin{figure*}
    \centering
    \includegraphics[height=0.32\linewidth]{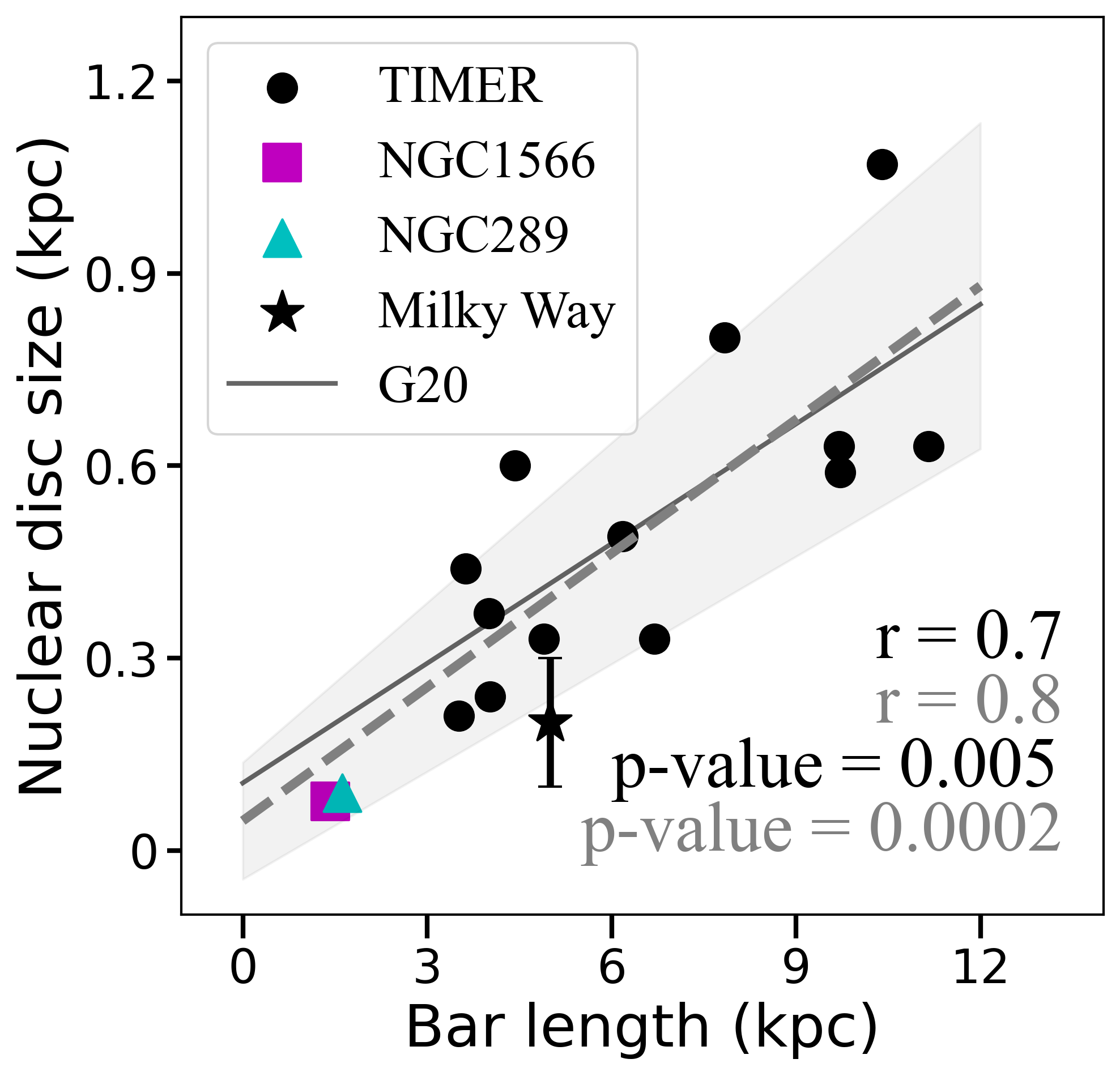}
    \includegraphics[height=0.32\linewidth]{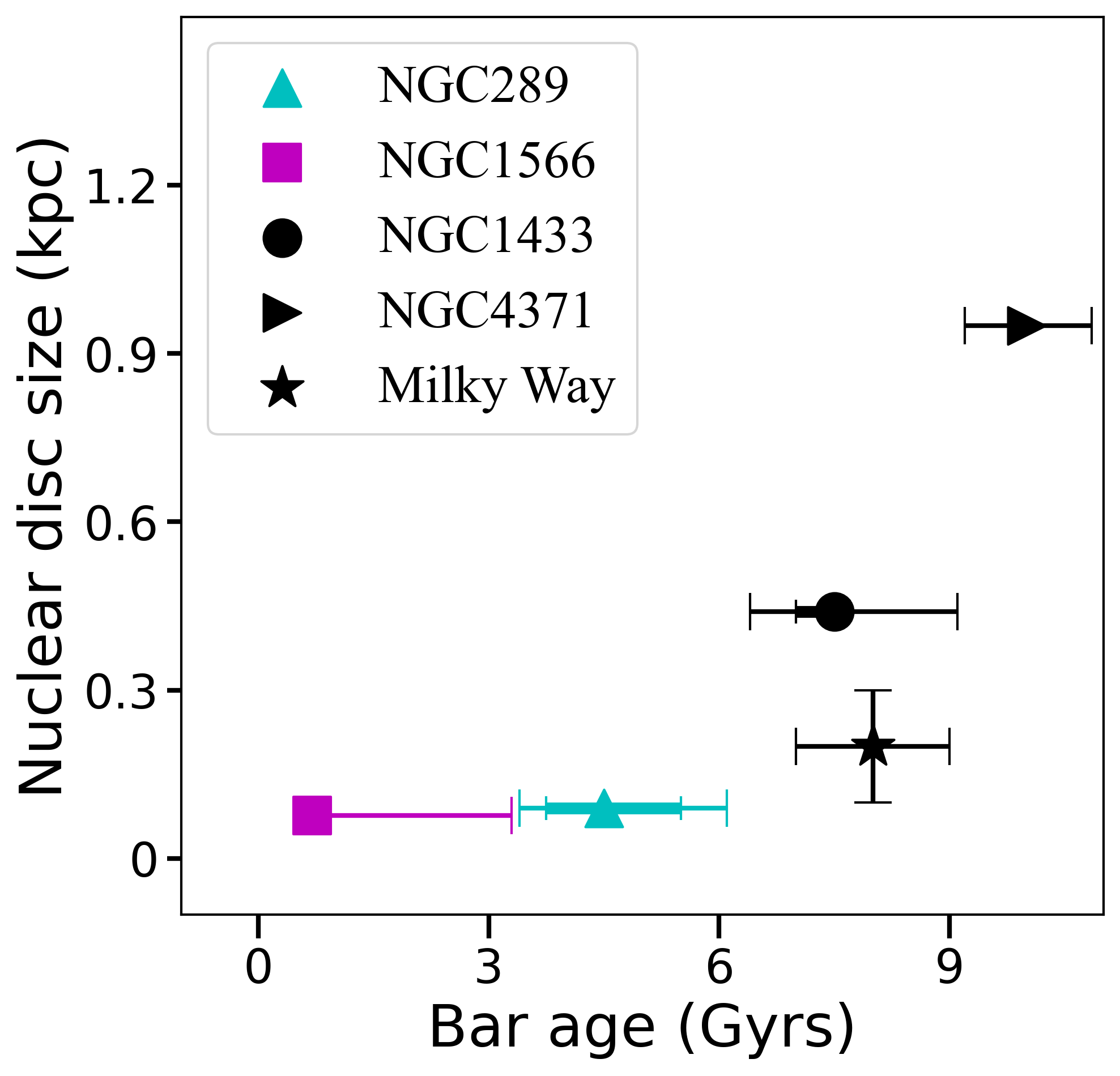}
    \includegraphics[height=0.32\linewidth]{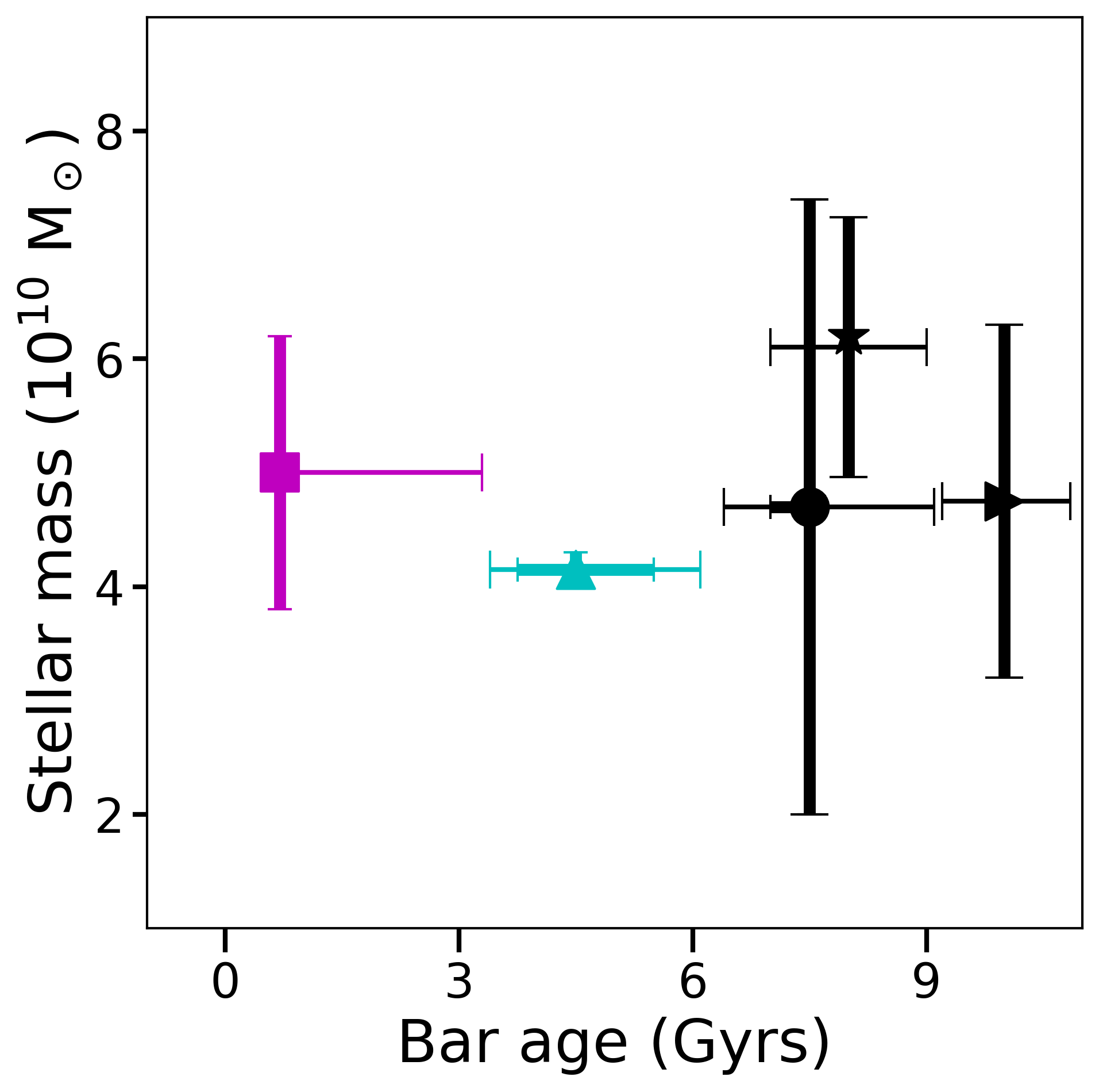}
    \caption{\textit{\textbf{The smallest nuclear disc and their young bars in context}.} \textbf{In the left}, we show the relation of nuclear disc size with bar length from the TIMER sample (\citealp{gadotti2020kinematic} -- black circles), together with the two galaxies from this paper -- NGC\,289 (cyan triangle), and NGC\,1566 (magenta square) -- and values for the Milky Way (black star). For the nuclear disc size for the Milky Way, we consider $\sim 100-300~\rm{pc}$ (\citealp{sormani2020jeans},\citeyear{sormani2022self}) and for the bar length, $5.0\pm0.2$ kpc (\citealp{wegg2015structure}). We also display the linear regression for the TIMER sample alone (solid black line) and considering this work, with the two new galaxies (dashed gray line). With the galaxies in this work, the Pearson correlation coefficient between nuclear disc size and bar length is strengthened from $0.73$ (TIMER only) to $0.82$ (this work). We do not consider the Milky Way for linear regression. The two galaxies from this work host considerably smaller nuclear discs than the ones in the TIMER sample. \textbf{In the center}, we show the relation of nuclear disc size with bar age. We consider the values for NGC\,1433 (\citealp{de2023new}), NGC\,4371 (\citealp{gadotti2015muse}, \citeyear{gadotti2020kinematic}), the Milky Way (\citealp{sormani2020jeans}, \citeyear{sormani2022self}; \citealp{wylie2022milky}; \citealp{sanders2022mira}) and the two galaxies from this work, NGC\,289 and NGC\,1566. The error bars of NGC\,289 and NGC\,1566 are the statistical and systematic errors, measured in this work; for NGC\,1433, we considered the statistical and systematic errors from \cite{de2023new}; for the Milky Way, we considered the different values from the literature, and for NGC\,4371, we considered the measured errors from \cite{gadotti2015muse}. It is clear that the bar ages measured for NGC\,289 and NGC\,1566 are the youngest, even when considering the error bars. \textbf{In the right}, we display the values of the total stellar mass as a function of bar age. For stellar mass values, we consider the mean value of different literature references (see Table \ref{tab_stellarMasses}). We do not consider the extrapolated values for total stellar mass from this work. With the information from the $5$ galaxies, we find no correlation. This could indicate that downsizing is not sufficient to determine bar formation, although more data is needed to achieve robust results.}
    \label{fig_TIMER}
\end{figure*}

In this work, we report the smallest kinematically confirmed nuclear discs, as well as the youngest bar, ever discovered, to the best of our knowledge. The nuclear discs hosted by NGC\,289 and NGC\,1566 have respective sizes of $90^{+16}_{-8}$ and $77^{+47}_{-2}$ pc. In Sect. \ref{sec_evidenceSmallNuclearDisc} we present the characteristics of each nuclear disc, together with spatial maps of the kinematic and stellar population properties. Both galaxies show all the kinematic characteristics expected for the presence of a second ordered fast-rotating structure in the centre of the galaxy, the nuclear disc. Additionally, the stellar population properties of both galaxies are those expected if the nuclear disc is formed by gas infall due to the presence of the bar. When compared to the surroundings, both nuclear discs have younger mean stellar ages and lower [$\alpha$/Fe] values. These properties indicate the late formation of the nuclear disc when compared to the main disc that was already present. Also, the lower values of [$\alpha$/Fe] indicate a continuous star formation, the opposite of a sudden starburst driven by mergers, which would present higher values of [$\alpha$/Fe] enhancement. Additionally, NGC\,289 presents higher values of [M/H] in the nuclear disc when compared to the surroundings, which is also expected in the bar-built scenario. Lastly, on the contrary, NGC\,1566 presents lower values of metallicity, which is not expected for the bar-built scenario of the nuclear disc. However, \cite{gadotti2019time} report a similar case in the TIMER survey, NGC\,1097, in which the nuclear ring also presents low values of [M/H], indicating the lack of pre-processing. As in NGC\,1097, NGC\,1566 has signs of a recent interaction and a low-mass satellite companion, NGC\,1581, which could explain the origin of the low-metalicity gas. Also, \cite{bittner2020inside} find similar radial trends for $3$ nuclear discs and $8$ nuclear rings out of $17$ galaxies. Since the metallicity values are intrinsically connected to the origin and history of the gas brought inwards -- which we do not know and is beyond the scope of this work -- it is possible that the original gas was not enriched for unknown reasons, and can still be in agreement with the bar-built scenario of nuclear discs. 

When compared to known nuclear discs, such as the ones reported by the TIMER collaboration (left panel in Fig. \ref{fig_TIMER} -- \citealp{gadotti2019time}), the nuclear discs we find are the smallest reported. Interestingly, both galaxies also have smaller bars than the TIMER sample, with sizes comparable to nuclear bars (\citealp{erwin2004double}). However, there is no evidence of longer bars in both cases, and we therefore consider these to be the main bar of the galaxy, albeit with lengths that are at the low end of the observed distribution. In fact, \cite{erwin2005large} finds that the mean main bar size in late-type disc galaxies (Sc-Sd) is $1.5~\rm{kpc}$, very close to the bars in both galaxies studied here (although note that both are Sbc galaxies). This makes them interesting objects to investigate further the evolutionary link between bars and nuclear discs. There is growing evidence that supports that bars and nuclear discs can evolve simultaneously, both from simulations and observations (e.g., \citealp{shlosman1989bars}; \citealp{knapen2005structure}; \citealp{comeron2010ainur}; \citealp{seo2019effects}; \citealp{gadotti2020kinematic}). From observations, \cite{gadotti2020kinematic} show a clear relation between the kinematic nuclear disc size and the bar length (see left panel in Fig. \ref{fig_TIMER}), which can imply a possible co-evolution. Furthermore, studies indicate that the nuclear disc grows inside-out with time (e.g., \citealp{bittner2020inside}; \citealp{de2023new}), in agreement with simulations (\citealp{seo2019effects}). In fact, \cite{seo2019effects} predict that, for Milky-Way galaxies, bar-built nuclear discs can form as small as $40$ pc -- depending on properties such as gas fraction and dynamics --, which is in agreement with our findings. Adding the two galaxies from this work, the correlation between nuclear disc size and bar length is strengthened from a Person-coefficient of $0.72$ to $0.83$, with a \textit{p-value} of $2\times10^{-4}$ (see Fig. \ref{fig_TIMER}, left panel). This is consistent with the scenario in which the nuclear disc growth is connected to the bar length, although the correlation itself does not necessarily imply causality. Nonetheless, exactly which mechanisms are responsible for defining the size of the nuclear disc and how are still debated (see, e.g., \citealp{sormani2018dynamical}).  

Finally, in this work, we find the first kinematically confirmed extragalactic nuclear disc as small as the one in our Galaxy, which has a size of $100-200$ pc (e.g., \citealp{launhardt2002nuclear}). The differences in the size of the nuclear disc from the Milky Way to extragalactic can be either real, in the sense that our Galaxy hosts a small nuclear disc, or artificial, due to different measurement methods. Although we cannot rule out that the differences can arise from different measurement methods, our findings show that small nuclear discs ($\sim 100$ pc) exist and can be found in other galaxies as well.


\subsection{Bars are still forming and discs are still settling}
\label{subSec_Disc_Bars}

A number of theoretical and observational studies find that bars are robust, long-lived structures, and once formed, cannot be easily destroyed (e.g., \citealp{athanassoula2003angular}; \citealp{athanassoula2005can}; \citealp{kraljic2012two}; \citealp{gadotti2015muse}; \citealp{perez2017observational}; \citealp{de2019clocking}; \citealp{rosas2020buildup}; \citealp{fragkoudi2020chemodynamics}, \citeyear{fragkoudi2021revisiting}; \citealp{de2023new}). Additionally, by studying how the fraction of barred galaxies evolves with redshift, it becomes clear that bars exist at least since $z \leq 1 - 2$ (e.g., \citealp{sheth2008evolution}; \citealp{melvin2014galaxy}; \citealp{guo2023first}), and the fraction increases with time (e.g., \citealp{sheth2008evolution}; \citealp{cameron2010bars}; \citealp{melvin2014galaxy}; \citealp{zhao2020barred}). In fact, in the Local Universe, bars are common structures and are present in $30\%-70\%$ of the galaxies (e.g., \citealp{eskridge2000frequency}; \citealp{menendez2007near}; \citealp{barazza2008bars}; \citealp{aguerri2009population}; \citealp{nair2010fraction}; \citealp{buta2015classical}; \citealp{erwin2018dependence}). Although it is not completely clear what are the fundamental properties of galaxies to lead them to form and sustain a bar, analytical and numerical works indicate that the moment of bar formation is linked to the dynamical settlement of the disc (e.g., \citealp{kraljic2012two}). That is, galaxies can only form and sustain a bar once their discs are massive enough and sufficiently dynamically cold, at least partially. Due to that, massive galaxies are expected to achieve a minimum mass to settle first, following the downsizing scenario (e.g., \citealp{cowie1996new}; \citealp{thomas2010environment}; \citealp{sheth2012hot}). In that scenario, one could expect a relation between galaxy mass and bar age, where the oldest bars would be found in massive galaxies and, on the other hand, young bars in less massive galaxies.

Following the methodology presented in \cite{de2023new}, we measure the bar ages of $4.50^{+1.60}_{-1.10}\rm{(sys)}^{+1.00}_{-0.75}\rm{(stat)}$ and $0.7^{+2.60}\rm{(sys)}^{+0.05}_{-0.05}\rm{(stat)}$ Gyr for NGC\,289 and NGC\,1566, respectively. To the best of our knowledge, these are the youngest bars for which we have a robust estimate of their ages. Since bar formation is associated with disc settling, our findings indicate that the discs in these galaxies recently settled or are still partially settling. Additionally, analyzing photometric and kinematic properties following \cite{erwin2017frequency} and \citeauthor{mendez2008confirmation} (\citeyear{mendez2008confirmation}, \citeyear{mendez2019inner}), respectively, we did not find any evidence of a presence of a box/peanut bulge. More specifically, we looked for `spurs' signatures in S$^4$G images and analyzed the $h_4$ along the bar major axis. For both analyses, we did not find signs of the presence of a box/peanut bulge. For more details regarding the photometric analysis, we refer the reader to \cite{erwin2017frequency} and, for the kinematic analysis, to \citeauthor{mendez2008confirmation} (\citeyear{mendez2008confirmation}, \citeyear{mendez2019inner}). Since it is expected that bars take $\sim3-4$ Gyr to develop a box/peanut bulge (e.g., \citealp{perez2017observational}), this is in line with the fact that these bars are young and recently formed. Considering the scenario in which bars form nuclear discs, it is not surprising that the smallest nuclear discs are hosted by young bars. In fact, this is expected in the bar-driven and inside-out growth scenarios. That is, recently formed bars would host small nuclear discs (e.g., \citealp{seo2019effects}). In summary, the measured bar ages together with the nuclear disc sizes from our work support scenarios of co-evolution between the bar and nuclear disc, and the inside-out growth of the nuclear disc itself, even if provisionally.

Analyzing our sample in the context with other findings, for the first time we can start to investigate the relationship between nuclear disc size and bar age (see Table \ref{tab_barAges} and Fig. \ref{fig_TIMER} -- middle panel), which will allow us to understand how nuclear discs grow in size in the future once we derive more bar ages using the full TIMER sample. Considering the current sample of galaxies for which we do have the measured bar age -- NGC\,1433 ($7.5$~Gyr -- \citealp{de2023new}); NGC\,4371 ($10$~Gyr -- \citealp{gadotti2015muse}); and the Milky Way ($8$~Gyr -- \citealp{wylie2022milky}; \citealp{sanders2022mira}), in addition to NGC 289 and NGC 1566 --, we show a tentative exponential growth scenario for nuclear discs in galaxies with similar stellar masses (see Table \ref{tab_stellarMasses}). Whereupon the nuclear disc, at first, hardly shows a development until $\sim6$~Gyr, followed by fast growth. Despite our small sample, this scenario is in qualitative agreement with theoretical expectations (e.g. \citealp{seo2019effects}), that find that young nuclear discs form small and are repetitively destroyed by their own star formation. This can also explain the lack of small nuclear discs discovered. Once they accumulate enough mass and the bar grows long enough, the nuclear disc can effectively grow. However, \cite{seo2019effects} find this transition to happen after $\sim2$~Gyr, depending on the simulation configurations. Nevertheless, this is a preliminary result and a larger sample is needed to robustly understand how nuclear discs grow, which is one of the TIMER collaboration goals for the future. In addition, more simulations are needed, in particular simulations employing a cosmological setting, to better understand the formation and growth of nuclear discs. 

\renewcommand{\arraystretch}{1.6}
\begin{table*}[]
\centering
\begin{tabular}{ccc|cc}
\textbf{Galaxy} & \textbf{Bar length (Kpc)}  & \textbf{Nuclear disc size (Kpc)} & \textbf{Bar age (Gyr)} & \textbf{Reference} \\
\hline
\hline
NGC\,289 & $1.62$ & $0.090$ & $4.5^{+1.60}_{-1.10}\rm{(sys)}^{+1.00}_{-0.75}\rm{(stat)}$ & This work \\
NGC\,1566 & $1.40$ & $0.077$ & $0.7^{+2.60}\rm{(sys)}^{+0.05}_{-0.05}\rm{(stat)}$ & This work \\
NGC\,1433 & $3.63$ & $0.380$ & $7.5^{+1.60}_{-1.10}\rm{(sys)}^{+0.2}_{-0.5}\rm{(stat)}$ & \cite{de2023new} \\
NGC\,4371 & $5.20$ & $0.950$ & $10.0\pm0.8$ & \cite{gadotti2015muse} \\
Milky Way & $5.00$ & $0.200$ & $8.0$ & \cite{wylie2022milky}, \cite{sanders2022mira} \\
\end{tabular}
\caption{Properties of galaxies considered in Fig. \ref{fig_TIMER} derived from different studies. The bar lengths for NGC\,289 and NGC\,1566 are from \cite{munoz2015spitzer}, NGC\,1433 from \cite{kim2014unveiling}, and NGC\,4371 from \cite{herrera2015catalogue}. The kinematic nuclear disc sizes for NGC\,1433 and NGC\,4371 are from \cite{gadotti2020kinematic}. Finally, the nuclear disc size and bar length values of the Milky Way are from \cite{launhardt2002nuclear} and \cite{wegg2015structure}.}
\label{tab_barAges}
\end{table*}

Finally, we also investigate how our sample fits the downsizing scenario (see Fig. \ref{fig_TIMER} -- right panel), in which massive galaxies are expected to host older bars. To investigate this scenario, we considered different mass measurements in the literature, summarized in Table \ref{tab_stellarMasses}. Contrary to the expected, our galaxies that host young bars have similar stellar masses of galaxies with bars as old as $10$~Gyr. This indicates that the downsizing scenario may not be sufficient to explain bar formation, but other processes may also be needed. In other words, even if galaxies have enough mass, other factors can be limiting bar formation, and further investigation is needed. In fact, bars can also form due to tidal interactions, including interactions with satellite galaxies, and this mechanism of bar formation may be independent of the galaxy mass (see, e.g., \citealp{noguchi1987close}; \citealp{gerin1990influence}; \citealp{lokas2021bar}). This could be the case of NGC\,289 and NGC\,1566 since both galaxies have close companions and signs of recent interactions. Nevertheless, from Table \ref{tab_stellarMasses}, it is clear that stellar mass measurement is not trivial, and different methods can result in masses differing by a factor $\sim3$. Additionally, we present here a small sample and a tentative result. In the near future, we will analyze the same properties for the entire TIMER sample (\citealp{gadotti2019time}), which will enable us to derive a more robust scenario.

\section{Summary and conclusions}
\label{sec_Summary}

In this work, we report the smallest kinematically confirmed nuclear discs observed to date. Additionally, applying the methodology from \cite{de2023new}, we measure their respective bar ages, and find that their bars are also the youngest bars to date for which there are bar age estimates. We summarize our findings as follows:

\begin{itemize}
    \item We report evidence for the serendipitous discovery of nuclear discs with sizes of $90^{+16}_{-8}$ and $77^{+47}_{-2}$ pc in NGC\,289 and NGC\,1566, respectively. We analyzed spatially resolved kinematic and stellar population properties for both galaxies. Both galaxies present all the kinematic characteristics of a secondary fast-rotating central structure, the nuclear disc. In addition, their nuclear discs present most of the average stellar population properties expected for a bar-driven formation. These properties follow the scenario in which the nuclear disc is formed by gas inflow triggered by the bar formation (Section \ref{sec_evidenceSmallNuclearDisc}).
    
    \item We measured the ages for both bars hosting the nuclear discs (\citealp{de2023new}) and find ages of $4.50^{+1.60}_{-1.10}\rm{(sys)}^{+1.00}_{-0.75}\rm{(stat)}$ and $0.7^{+2.60}\rm{(sys)}^{+0.05}_{-0.05}\rm{(stat)}$ Gyr for NGC\,289 and NGC\,1566, respectively. This is in agreement with the bar-driven and inside-out growth scenarios, in which young bars form small nuclear discs, and, as the bar grows longer, the nuclear disc grows larger (Section \ref{sec_timingBarFormation}).
    
    \item Analyzing the bar length and nuclear disc size relation together with the TIMER sample (\citealp{gadotti2019time}), we find that our sample agrees with the correlation. In fact, by adding our two galaxies, the correlation is strengthened from  $\rm{r} = 0.73$ to $0.82$ with a \textit{p-value} of $2\times10^{-4}$. This is in agreement with the nuclear disc growing inside out with time (Section \ref{subSec_SmallestNuclearDisc}).
    
    \item Analyzing the bar age with nuclear disc size relation, together with $3$ galaxies from the literature (NGC\,1433, NGC\,4371, and the Milky Way), we find a suggestive exponential relation. In that scenario, nuclear discs would take longer to effectively grow. This is in qualitative agreement with theoretical works (e.g. \citealp{seo2019effects}) which suggest that nuclear discs grow in time (Section \ref{subSec_Disc_Bars}).
    
    \item Analyzing the bar age with the galaxy stellar mass relation together with $3$ galaxies from the literature (NGC\,1433, NGC\,4371, and the Milky Way), we do not find a correlation between the bar age with the galaxy stellar mass. Although this finding might challenge the downsizing scenario for bar formation, whereby more massive galaxies would host older stellar bars, we also point out that our sample size is still rather limited. Nevertheless, we emphasize that measuring stellar mass is not trivial, and different methods find different masses (Section \ref{subSec_Disc_Bars}).
\end{itemize}

These results provide further intriguing evidence of the interplay between nuclear discs and the formation and evolution of bars. By applying the methodology developed in \cite{de2023new} to the entire TIMER sample (\citealp{gadotti2019time}) we will be able to increase our sample size, thus enabling us to probe the role played by downsizing on bar formation and the intricate interplay between bar formation and nuclear disc evolution.



\begin{acknowledgements}
      We thank the anonymous referee for the insightful comments. Raw and reduced data are available at the ESO Science Archive Facility. This work was supported by STFC grant ST/T000244/1 and ST/X001075/1. AdLC acknowledges support from Ministerio de Ciencia e Innovación through the Spanish State Agency (MCIN/AEI) and from the European Regional Development Fund (ERDF) under grant CoBEARD (reference PID2021-128131NB-I00), and under the Severo Ochoa Centres of Excellence Programme 2020-2023 (CEX2019-000920-S). IMN and JFB acknowledge support from grant PID2019-107427GB-C32 from the Spanish Ministry of Science and Innovation and from grant ProID2021010080 and CEX2019-000920-S in the framework of Proyectos de I+D por organismos de investigación y empresas en las áreas prioritarias de la estrategia de especialización inteligente de Canarias (RIS-3). MQ acknowledges support from the Spanish grant PID2019-106027GA-C44, funded by MCIN/AEI/10.13039/501100011033. PC acknowledges support from Conselho Nacional de Desenvolvimento Cient\'ifico e Tecnol\'ogico (CNPq) under grant 310555/2021-3 and from Funda\c{c}\~{a}o de Amparo \`{a} Pesquisa do Estado de S\~{a}o Paulo (FAPESP) process number 2021/08813-7. JN acknowledges funding from the European Research Council (ERC) under the European Union’s Horizon 2020 research and innovation programme (grant agreement No. 694343). PSB acknowledges the support of the Spanish Ministry of Science and Innovation through grant  PID2019-107427GB-C31.

\end{acknowledgements}

%
\bibliographystyle{aa} 
\bibliography{bibliography} 
%

\end{document}